\documentclass[12pt]{article}
\def\slashchar#1{\setbox0=\hbox{$#1$}           
   \dimen0=\wd0                                 
   \setbox1=\hbox{/} \dimen1=\wd1               
   \ifdim\dimen0>\dimen1                        
      \rlap{\hbox to \dimen0{\hfil/\hfil}}      
      #1                                        
   \else                                        
      \rlap{\hbox to \dimen1{\hfil$#1$\hfil}}   
      /                                         
   \fi}                                         %
\newcommand{\blank}{}
\renewcommand{\theequation}{\blank \arabic{equation}}
\newcounter{dummy}{}
\newcommand{\letters}{%
    \setcounter{dummy}{\value{equation}}
    \renewcommand{\thedummy}{\blank \arabic{dummy}}
    \renewcommand{\theequation}{\thedummy\alph{equation}}
    \refstepcounter{dummy}
    \setcounter{equation}{0}%
 }
\newcommand{\noletters}{%
    \setcounter{equation}{\value{dummy}}
    \renewcommand{\theequation}{\blank\arabic{equation}}%
  }
\newenvironment{mathletters}{\letters}{\noletters}
\newcommand{\dirac}{-i\slashchar{\nabla}}

\begin{document}
\title{Spectral Boundary Conditions in the  Bag Model}
\author{A.~A.~Abrikosov,~jr.\thanks{E-mail:
{\sc persik@itep.ru}} \\
{\em 117 259, Moscow, ITEP.}}
\date{}
\maketitle
\begin{abstract}
We propose a reduced form of spectral boundary conditions for
holding fermions in the bag in a chiral invariant way. Our
boundary conditions do not depend on time and allow Hamiltonian
treatment of the system. They are suited for studies of chiral
phenomena both in Minkowski and Euclidean spaces.

\end{abstract}

\section*{Introduction}

The two principal problems of QCD are confinement and
spontaneous breaking of chiral invariance. Both phenomena take
place in the strongly interacting domain where the theory
becomes nonperturbative. Probably they are interrelated.
However, usually they were considered separately. Up to now the
spontaneous chiral invariance breaking (SCIB) was discussed
mostly in the infinite space. It would be interesting to address
the features of SCIB that appear due to localization of quarks
in finite volume. In order to do that one needs to hold quarks
in a chiral invariant way.

There exists an entire family of bag models. The famous MIT bag
\cite{MIT} successfully reproduced the spectrum and other features
of hadrons. A generalization of the MIT model are so-called chiral
bags \cite{wipf/duerr,esposito/kirsten}. An apparent drawback of
all these models is that the boundary conditions are explicitly
chiral noninvariant.

Attempts to save the situation led to the cloudy bag
\cite{theberge} where the chiral symmetry was restored by pions
emitted from the bag surface (the pion cloud). But this model is
sensitive to details of the adopted scheme of quark-pion
interaction. Thus neither of the models is suited to the
discussion of SCIB in finite volume.

The way to lock fermions in a finite volume without spoiling the
chiral invariance is to impose the so-called \textbf{spectral
boundary conditions} (SBC). They were first introduced by Atiah,
Patodi and Singer (APS) who investigated anomalies on manifolds
with boundaries \cite{APS}. Later these boundary conditions were
widely applied in studies of index theorems on various manifolds
\cite{euguchi/gilkey/hanson}.

In distinction from those mentioned before the APS conditions are
nonlocal. They are defined on the boundary as a whole. This looks
natural for finite Euclidean manifolds but is inconvenient for
physical models. In the process of evolution the spatial boundary
of a static physical bag becomes an infinite space-time cylinder.
Constraining fields on the entire ``world cylinder'' would violate
causality and complicate continuation to Minkowski space.

In this talk we shall demonstrate how to avoid this difficulty.
One can consider a restricted spatial version of spectral
boundary conditions. The modified conditions do not depend on
time and are acceptable from the physical point of view. This
makes possible the Hamiltonian description of the system.

The paper has the following structure. We shall review the
classical APS boundary conditions in Section~1. In Section~2 we
shall formulate the modified spectral conditions and discuss their
properties. In conclusion we shall summarize the results and
mention future prospects.

\section{The physics of APS boundary conditions}

\subsection{Conventions}

We shall start from the traditional form of SBC. First we will
introduce coordinates, Dirac matrices and the gauge that allow to
define the spectral boundary conditions in a convenient way.

Let us consider massless fermions interacting with gauge field
$\hat{A}$ in a closed 4-dimensional Euclidean domain $B_4$. We
choose a curvilinear coordinate frame such that in the vicinity of
the boundary $\partial B_4$ the first coordinate $\xi$ points
along the outward normal ($\xi=0$ corresponds to $\partial B_4$)
while the three others, $q^i$, parametrize $\partial B_4$ itself.
For simplicity we shall assume that near the surface the metric
 $g_{\alpha \beta}$ depends only on $q$ so that
\begin{equation}\label{g-APS}
  ds^2 =  d \xi^2 + g_{ik}(q)\, dq^i\, dq^k .
\end{equation}
Moreover we choose the gauge so that on the boundary
 $\hat{A}_\xi = 0$.

Now we must fix the Dirac matrix $\gamma^\xi$. Let $I$ be the
$2\times 2$ unity matrix. Then
\begin{equation}\label{gamma-APS}
  \gamma^\xi = \left( \begin{array}{cc}
  0         & iI \\
  -iI & 0
\end{array} \right);
\qquad \qquad
  \gamma^q = \left( \begin{array}{cc}
  0         & \sigma^q \\
  \sigma^q  & 0
\end{array} \right).
\end{equation}
Matrices $\sigma^q$ are the ordinary Pauli $\sigma$-matrices. With
these definitions the Dirac operator of massless fermions on the
surface takes the form,
\begin{equation}\label{nabla-APS}
    \left.
        -i \slashchar{\nabla}
    \right|_{\partial B_4} =
         -i \gamma^\alpha \nabla_\alpha =
 \left(\begin{array}{cc}
  0                 & \hat{M}  \\
  \hat{M}^\dagger   & 0
\end{array} \right) =
 \left(\begin{array}{cc}
  0                 & I \, \partial_\xi - i \hat{\nabla}  \\
   - I \, \partial_\xi - i \hat{\nabla} & 0
\end{array} \right) ,
\end{equation}
where $\hat{\nabla} = \sigma^q\, \nabla_q$ is the convolution of
covariant gradient along the boundary $\nabla_q$ with
$\sigma$-matrices. Note that Hermitian conjugated operators
$\hat{M}$ and $\hat{M}^\dagger$ differ only by the sign of
$\partial_\xi$-derivative.

Further on we shall call the covariant derivative
 $-i \hat{\nabla}$ on the boundary the \textbf{boundary operator.} It
is a linear differential operator acting on 2-spinors. It is
Hermitian and includes tangential gauge field $\hat{A}_q$ and
the spin connection which arises from the curvature of $\partial
B_4$.

The massless Dirac operator anticommutes with $\gamma^5$-matrix:
\begin{equation}\label{chirality}
  \left \{-i \slashchar{\nabla},\, \gamma^5 \right\} = 0,
\qquad
        \gamma^5 = \left(
\begin{array}{rr}
  I & 0 \\
  0 & -I
\end{array} \right).
\end{equation}
Thus the Lagrangian of Quantum Chromodynamics is chiral invariant.
In order to exploit this one needs chiral invariant boundary
conditions.

\subsection{The APS boundary conditions} \label{APS}
\subsubsection{The definition}

Atiah, Patodi and Singer investigated spectra of Dirac operator on
manifolds with boundaries. If we separate upper and lower (left
and right) components of 4-spinors the corresponding eigenvalue
equation for $-i\slashchar{\nabla}$ will take the form
\begin{equation}\label{eigenvalue}
 -i\slashchar{\nabla}\, \psi_\Lambda =
 -i\slashchar{\nabla} \left(
\begin{array}{c}
u_\Lambda  \\ v_\Lambda
\end{array}\right) =
 \Lambda \left(
\begin{array}{c}
u_\Lambda  \\ v_\Lambda
\end{array}\right ) =
 \Lambda\, \psi_\Lambda.
\end{equation}
The next step is to Fourier expand $u$ and $v$ on the boundary.
Let 2-spinors $e_\lambda (q)$ be eigenfunctions of the boundary
operator $-i\hat{\nabla}$:
\begin{equation}\label{e-lambda}
  -i\hat{\nabla}\, e_\lambda (q) = \lambda\, e_\lambda (q).
\end{equation}
Note that the form of this equation and the eigenfunctions
$e_\lambda (q)$ depend on gauge. It is here that the gauge
condition
 $\hat{A}_\xi (0,\, q) = 0$ becomes important.

The operator $-i\hat{\nabla}$ is Hermitian so $\lambda$ are real.
The functions $e_\lambda$ form an orthogonal basis. In principle
$-i\hat{\nabla}$ may have zero-modes on $\partial B_4$ but convex
manifolds are not the case.

In the vicinity of the boundary spinors $u_\Lambda$ and
$v_\Lambda$ may be expanded in series in $e_\lambda$:
\begin{mathletters}\label{uv-exp}
\begin{eqnarray}
  u_\Lambda (\xi,\, q) & = &
    \sum_\lambda f_\Lambda^\lambda (\xi)\, e_\lambda (q),
\quad
    f_\Lambda^\lambda (\xi)
        = \int_{\partial B_4}
        e_\lambda^\dagger (q)\, u_\Lambda (\xi,\, q)\,
        \sqrt g\, d^3 q;
\label{uv-exp/a} \\
  v_\Lambda (\xi,\, q) & = &
    \sum_\lambda g_\Lambda^\lambda (\xi)\, e_\lambda (q),
\quad
    g_\Lambda^\lambda (\xi)
        = \int_{\partial B_4}
        e_\lambda^\dagger (q)\, v_\Lambda (\xi,\, q)\,
        \sqrt g\, d^3 q;
\label{uv-exp/b}
\end{eqnarray}
\end{mathletters}
where $g=\det ||g_{ik}||$ is the determinant of metric on the
boundary.

The spectral boundary condition states that on the boundary,
\emph{i.~e.\/} at $\xi =0$
\begin{mathletters}\label{SBC}
\begin{eqnarray}
  \left.
    f_\Lambda^\lambda
  \right|_{\partial B_4} & = & 0
    \qquad \mathrm{for} \qquad \lambda > 0;  \label{SBC/a} \\
  \left.
    g_\Lambda^\lambda
  \right |_{\partial B_4} & = & 0
    \qquad \mathrm{for} \qquad \lambda < 0.  \label{SBC/b}
\end{eqnarray}
\end{mathletters}
Another way to say this is to introduce integral projectors
$\mathcal{P}^+$ and $\mathcal{P^-}$ onto boundary modes with
positive and negative $\lambda$:
\begin{equation}\label{Projectors}
  \mathcal{P}^+ (q,\, q') =
    \sum_{\lambda > 0} e_\lambda (q) \, e_\lambda^\dagger (q');
\qquad
  \mathcal{P}^- (q,\, q') =
    \sum_{\lambda < 0} e_\lambda (q) \, e_\lambda^\dagger (q').
\end{equation}
Let $\mathcal{I}$ be the unity operator on the function space
spanned by $e_\lambda$. Then, obviously,
\begin{equation}\label{Ppls/Pmns}
    \mathcal{P^+ + P^- = I .}
\end{equation}
If we join two-dimensional projectors $\mathcal{P}^+$ and
$\mathcal{P^-}$ into $4\times 4$ matrix $\mathcal{P}$ the spectral
boundary condition  for 4-spinor $\psi$ will look as follows:
\begin{equation}\label{P-APS}
    \left.
        \mathcal{P}\, \psi
    \right|_{\partial B_4} =
    \left.
    \left(
        \begin{array}{cc}
          \mathcal{P}^+ & 0 \\
          0 & \mathcal{P}^- \\
        \end{array}
    \right)
    \left(
        \begin{array}{c}
          u \\
          v \\
        \end{array}
    \right)
    \right|_{\partial B_4} = 0.
\end{equation}
The projector $\mathcal{P}$ commutes with matrix $\gamma^5$:
\begin{equation}\label{[P,gamma5]}
    \left[
        \mathcal{P},\, \gamma^5
    \right] = 0.
\end{equation}
Therefore boundary condition (\ref{P-APS}) by construction
preserves the chiral invariance.

\subsubsection{The physics}

Now we shall prove that the spectral boundary conditions are
acceptable and explain their physical meaning. Namely, we shall
show that SBC provide Hermicity of the Dirac operator and
conservation of fermions in the bag. After that we will explain
the origin of requirements (\ref{SBC}).

First let us prove that Dirac operator is Hermitian. As usually,
we integrate by parts the expression
\begin{equation}\label{Dirac=Herm}
    \int_{B_4} dV\, f^\dagger\, (\dirac g) =
    \int_{B_4} dV\, (\dirac f)^\dagger\, g +
    \oint_{\partial B_4} dS\, f^\dagger\, (-i \gamma^\xi) g.
\end{equation}
Now we need to show that if $f$ and $g$ satisfy (\ref{SBC}) then
the last term vanishes.

Conditions (\ref{SBC}) mean that on the boundary 4-spinors $f$ and
$g$ may be written as:
    $f = \left(
        \begin{array}{c}
          f^- \\
          f^+ \\
        \end{array}
        \right)$
and
    $g = \left(
        \begin{array}{c}
          g^- \\
          g^+ \\
        \end{array}
        \right)$,
where $f^\pm$ and $g^\pm$ include only components with positive
and negative $\lambda$ respectively, see (\ref{uv-exp}). Rewriting
the boundary term in (\ref{Dirac=Herm}) we get
\begin{equation}\label{oint=0}
    \oint_{\partial B_4} dS\, f^\dagger\, (-i \gamma^\xi) g =
    \oint_{\partial B_4} dS\,
    \left[
        (f^-)^\dagger\, g^+ - (f^+)^\dagger\, g^-
    \right] = 0,
\end{equation}
due to the orthogonality of eigenfunctions of the boundary
operator. Thus the APS boundary conditions indeed ensure the
Hermicity of Dirac operator.

In addition, relation (\ref{oint=0}) guarantees conservation of
fermions in the bag. Indeed, for $f = g$ the LHS is nothing but
the net fermionic current through the boundary,
\begin{equation}\label{Fconserv}
    \oint_{\partial B_4} dS\, j^\xi =
    - i\oint_{\partial B_4} dS\, f^\dagger \gamma^\xi f = 0.
\end{equation}
Therefore the number of fermions is conserved and particles in the
spectral bag are confined.

In order to understand the origin of SBC let us rewrite the
eigenvalue condition (\ref{eigenvalue}) near the boundary in terms
of components.
\begin{mathletters}\label{fg-eqns}
\begin{eqnarray}
  (\partial_\xi + \lambda)\, g_\Lambda^\lambda (\xi) & = &
    \Lambda\, f_\Lambda^\lambda (\xi);
        \label{fg-eqns/a} \\
  - (\partial_\xi - \lambda)\, f_\Lambda^\lambda (\xi) & = &
    \Lambda\, g_\Lambda^\lambda (\xi).
        \label{fg-eqns/b}
\end{eqnarray}
\end{mathletters}
Depending on the sign of $\lambda$ these relations reduce on the
boundary either to
\begin{mathletters}\label{logD}
\begin{equation}\label{logD/a}
  \left. \frac{\partial_\xi g_\Lambda^\lambda }{g_\Lambda^\lambda }
    \right|_{\xi=0}
    = -\lambda <0 , \qquad f_\Lambda^\lambda (0) = 0
        \quad \mathrm{at} \quad \lambda > 0;
\end{equation}
or to
\begin{equation}\label{logD/b}
  \left. \frac{\partial_\xi f_\Lambda^\lambda }{f_\Lambda^\lambda }
    \right|_{\xi=0}
    = \hphantom{-} \lambda <0 , \qquad g_\Lambda^\lambda (0) = 0
        \quad \mathrm{at} \quad \lambda < 0.
\end{equation}
\end{mathletters}
Thus both components either vanish on the boundary or have a
negative logarithmic derivative along the normal.

This requirement has a simple physical interpretation. Suppose
that out of the bag the metric and the gauge field remain the same
as on the boundary. Then we can continue the functions $f$ and $g$
to $\xi = \infty$. Outside the bag the functions will be square
integrable falling exponents as if the particle was locked in a
potential well. The only difference is that now the potential for
every mode is adjusted specially. We may conclude that the
spectral boundary conditions claim that wave functions must have
square integrable continuation to the infinite space.

\section{The SBC for physical bags}

\subsection{The truncated SBC}

Now let us turn to fermions in the infinite Euclidean cylinder
$B_3\otimes R$. We shall call the first three coordinates
``space'' and the fourth one ``time''.  The boundary operator
consists of spatial and temporal parts:
\begin{equation}\label{BOfull}
    -i \hat{\nabla}_{\partial B_3 \otimes R} =
    -i \hat{\nabla}_{\partial B_3} - i \sigma^z \partial_4 .
\end{equation}
We shall call the spatial part
 $-i \hat{\nabla}_{\partial B_3}$ the
\textbf{truncated boundary operator}. Let its eigenfunctions be
$e^\pm_\lambda$:
\begin{equation}\label{e-lambda-3d}
  -i\hat{\nabla}_{\partial B_3}\, e^\pm_\lambda (q) =
  \pm\lambda\, e^\pm_\lambda (q),
\qquad
    \lambda > 0.
\end{equation}

Wave functions on the space-time boundary $\partial B_3\otimes R$
can be expanded in $e^\pm_\lambda$ and longitudinal plane waves:
\begin{mathletters} \label{Fourier+/-}
\begin{eqnarray}
  u_\Lambda  & = &
    \sum_{\lambda >0}
    \int \frac{dk}{2\pi}\,e^{ikt}\,
    \left[
        f^{+\lambda,\, k}_\Lambda \, e^+_\lambda +
        f^{-\lambda,\, k}_\Lambda \, e^-_\lambda
    \right] ;
\label{Fourier+/-a}\\
  v_\Lambda  & = &
    \sum_{\lambda >0}
    \int \frac{dk}{2\pi}\,e^{ikt}\,
    \left[
        g^{+\lambda,\, k}_\Lambda \, e^+_\lambda +
        g^{-\lambda,\, k}_\Lambda \, e^-_\lambda
    \right] .
\label{Fourier+/-b}
\end{eqnarray}
\end{mathletters}

The truncated operator $-i \hat{\nabla}_{\partial B_3}$
anticommutes with $\sigma^z$. Therefore $\sigma^z$ changes the
sign of $e$-eigenvalues. A possible choice of eigenvectors is (see
\cite{Leipzig01,abrikosov} for the sphere)
\begin{equation}\label{e+/e-}
  e^\pm_\lambda =
    \pm i \sigma^z \, e^\mp_\lambda .
\end{equation}
Thus the last term in (\ref{BOfull}) mixes positive and negative
spatial harmonics.

In classical approach this would mean that now SBC should be
written in terms of $k$-dependent eigenfunctions of the full
boundary operator (\ref{BOfull}). However physically these
``future-sensitive'' boundary conditions look strange. Therefore
we propose to apply independent of $k$ truncated APS constraints:
\begin{mathletters}\label{SBC3+1}
\begin{eqnarray}
    \left.
        f^{+\lambda,\, k}_\Lambda
    \right|_{\partial B_3} & = & 0 ;
\label{SBC3+1/a}
    \\
    \left.
        g^{-\lambda,\, k}_\Lambda
    \right|_{\partial B_3} & = & 0 .
\label{SBC3+1/b}
\end{eqnarray}
\end{mathletters}
These conditions do not depend on time and allow Hamiltonian
treatment of the system. Moreover, they may be applied both in
Euclidean and Minkowski spaces. Now let us show that they are
acceptable.

\subsection{Consistency}

We are going to prove that the truncated form of SBC fulfills
necessary conditions. Namely that they are chiral invariant, that
the Dirac operator is Hermitian and the fermionic current is
conserved and after all that wave functions may be continued out
of the bag.

The proof of the first three points literally follows the
4-dimensional case. Everything that concerns formulae
(\ref{Projectors}--\ref{Fconserv}) remains true for truncated
($_T$) 3-dimensional SBC (\ref{SBC3+1}). One may define on
$\partial B_3$ projectors,
\begin{equation}\label{3d-Projectors}
  \mathcal{P}^\pm_T (q,\, q') =
    \sum_{\lambda > 0} e^\pm_\lambda (q) \,
    \left[
        e^\pm_\lambda (q')
    \right]^\dagger.
\end{equation}
Then the truncated boundary conditions may be written in the
manifestly $\gamma^5$-invariant form,
\begin{equation}\label{3d-P-APS}
    \left.
        \mathcal{P}_T\, \psi
    \right|_{\partial B_3} =
    \left.
    \left(
        \begin{array}{cc}
          \mathcal{P}_T^+ & 0 \\
          0 & \mathcal{P}_T^- \\
        \end{array}
    \right)
    \left(
        \begin{array}{c}
          u \\
          v \\
        \end{array}
    \right)
    \right|_{\partial B_3} = 0.
\end{equation}

Hermicity of the Dirac operator and conservation of fermions are
proven in the same way as before, see
(\ref{Dirac=Herm}--\ref{Fconserv}). We don't rewrite the
formula.

The last point is more delicate. We already mentioned that the
$\sigma^z$ term in (\ref{BOfull}) mixes positive and negative
harmonics. Therefore they must be analysed together and instead of
two eigenvalue equations (\ref{fg-eqns}) we get four ($\xi$ is the
normal to the spatial boundary):
\begin{mathletters}\label{RHS3+1}
\begin{eqnarray}
      (\partial_\xi
      + \lambda)\,
      g_\Lambda^{+\lambda,\, k}\,
      & = &
    \Lambda\,  f_\Lambda^{+\lambda,\, k}
    + ik\,  g_\Lambda^{-\lambda,\, k};
\label{RHS3+1/a}
\\
    - (\partial_\xi
     - \lambda)\, f_\Lambda^{+\lambda,\, k}
     & = &
     \Lambda\, g_\Lambda^{+\lambda,\, k}
    + ik\, f_\Lambda^{-\lambda,\, k} :
\label{RHS3+1/b}
\\
      (\partial_\xi
      - \lambda)\,
      g_\Lambda^{-\lambda,\, k}\,
      & = &
    \Lambda\,  f_\Lambda^{-\lambda,\, k}
    - ik\,  g_\Lambda^{+\lambda,\, k};
\label{RHS3+1/c}
\\
    - (\partial_\xi
     + \lambda)\, f_\Lambda^{-\lambda,\, k}
     & = &
     \Lambda\, g_\Lambda^{-\lambda,\, k}
    - ik\, f_\Lambda^{+\lambda,\, k} .
\label{RHS3+1/d}
\end{eqnarray}
\end{mathletters}
The new feature with respect to (\ref{fg-eqns}) are $ik$ addends
that appear due to the mixing.  However one may notice that the
terms in the RHS of (\ref{RHS3+1}) come in pairs $f^+$, $g^-$ and
$f^-$, $g^+$. Therefore according to conditions (\ref{SBC3+1}) the
RHS of equations (\ref{RHS3+1/a}, \ref{RHS3+1/d}) vanish on the
boundary. Thus the behaviour of $g^+$ and $f^-$ on the boundary is
governed by the homogeneous equations and
\begin{equation}\label{logDf,g}
    \left.
        \frac{\partial_\xi f_\Lambda^{-\lambda,\, k} }%
        {f_\Lambda^{-\lambda,\, k}}
    \right|_{\xi=0} =
    \left.
        \frac{\partial_\xi g_\Lambda^{+\lambda,\, k} }%
        {g_\Lambda^{+\lambda,\, k}}
    \right|_{\xi=0} = -\lambda < 0.
\end{equation}

Hence despite the presence of extra pieces the nonvanishing
components $g^+$ and $f^-$ have negative logarithmic derivatives.
This means that solutions of eigenvalue equations may be continued
outwards of the ``world cylinder'' in an integrable way and the
last of our requirements is fulfilled. This completes the proof of
acceptability of the truncated SBC.

\section*{Conclusion}

The truncated version of APS boundary conditions offers a number
of possibilities. It allows to formulate a chiral invariant bag
model and to approach chiral properties of fermionic field in the
closed volume. The constraints are imposed on the spatial boundary
of the bag so one may write down the Hamiltonian and study the
energy spectrum of the system. Another advantage is that the
modified SBC do not depend on time and may be used both in
Euclidean and Minkowsky space.

A new feature that SBC may bring to the bag physics is their
nonlocality. Other bag models
\cite{MIT,wipf/duerr,esposito/kirsten} employed local boundary
conditions which correspond to the thin wall approximation. The
nonlocal spectral conditions refer to the boundary as a whole.
Therefore in a sense hadrons are also treated as a whole. This
looks promising by itself. To begin with it would be interesting
to investigate hadronic spectra in chiral invariant bags. This
could answer whether the model is realistic and indicate missing
elements.

Another question is more mathematical. Chiral symmetry is a
specific of fermions in even dimensional spaces. Hence the
spectral boundary conditions were also considered in even
dimensions. The truncated SBC are formulated in the odd
dimensional space that remains after discarding the time. This
might have interesting consequences. For example, the boundary of
odd-dimensional bag is an even-dimensional manyfold and one can
introduce a sort of internal chirality for surface modes. The
question is if there is a way for this hidden symmetry to reveal
itself.

In conclusion I would like to express my gratitude to Professor
A.~Wipf for elucidating discussions. I thank the Organizers for
financial support. The work was supported by RFBR grant
03--02--16209 and Federal Program of the Russian Ministry of
Industry, Science and Technology No 40.052.1.1.1112.


\end{document}